\documentclass[12pt]{article}
\pdfoutput=1
\usepackage{amsmath}
\usepackage{natbib}
\usepackage{amsfonts}
\usepackage{graphicx}
\usepackage{enumerate}
\usepackage{float}
\newcommand{\argmin}{\operatornamewithlimits{argmin}}

\newcommand{\blind}{0}

\begin{document}
	
\def\spacingset#1{\renewcommand{\baselinestretch}%
	{#1}\small\normalsize} \spacingset{1}


\if0\blind
{
	\title{\bf Minimizing the CDF Path Length: A Novel Perspective on Uniformity and Uncertainty of Bounded Distributions}
	\author{Michael E. Beyer, PE, CAP}

	\maketitle
} \fi

\if1\blind
{
	\bigskip
	\bigskip
	\bigskip
	\begin{center}
		{\LARGE\bf Title}
	\end{center}
	\medskip
} \fi

\bigskip
\begin{abstract}
	An index of uniformity is developed as an alternative to the maximum-entropy principle for selecting continuous, differentiable probability distributions $\mathcal{P}$ subject to constraints $C$. The uniformity index developed in this paper is motivated by the observation that among all differentiable probability distributions defined on a finite interval $[a,b] \in \mathbb{R}$, it is the uniform probability distribution that minimizes the \textit{path length} of the associated cumulative distribution function $F_{\mathcal{P}}$ on $[a,b]$. This intuition is extended to situations where there are constraints on the allowable probability distributions. In particular, constraints on the first and second raw moments of a distribution are discussed in detail, including the analytical form of the solutions and numerical studies of particular examples. The resulting "shortest path" distributions are found to be decidedly more heavy-tailed than the associated maximum-entropy distributions, suggesting that entropy and "CDF path length" measure two different aspects of uncertainty for bounded distributions.  
\end{abstract}

\noindent%
{\it Keywords:} Entropy, prior, uninformative, ignorance, Bayesian
\vfill

\newpage
\spacingset{1} 
\section{Introduction}
\label{sec:intro}
There are numerous situations were we seek a probability model to represent uncertainty, yet we only feel comfortable making a few assumptions about the nature of that uncertainty. In such situations, the choice of probability distributions is underdetermined, requiring additional criteria to pick out a single distribution from among those consistent with the assumed facts. A common criterion is the maximum entropy principle (MEP), which recommends selecting the distribution that maximizes the Shannon Entropy, which is defined for discrete probability distributions ($P_d$) as:

\[ H(X):=E_{P_d}[-\ln(P_d(X))]=-\sum\limits_{x_i \in X} P_d(x_i)\ln(P_d(x_i)) \]

For continuous distributions, one uses the Shannon Differential Entropy, which is an extension of the Shannon Entropy to probability density functions $f(x)$:

\[ h(X):=-\int_{X} f(x) \log(f(x)) dx\] 

The Shannon Entropy has several intuitively appealing properties when viewed as a measure of \textit{information content}: continuity, additivity, and maximality under a uniform distribution. It also has close connections to data processing and statistical mechanics, and it can be shown to satisfy several types of consistency criteria (\cite{Uffink96}). In his seminal paper introducing information theory \cite{Shannon} cited three properties that guided the development of the Shannon Entropy (H) for a discrete probability distribution $p_n(x_i)$, defined on a finite set $\{x_i\},\; i\in 1...n$:\\

\begin{enumerate}
	\item H should be continuous in the $p_n(x_i)$
	\item If all $p_n(x_i)$ are equal, then H should be a monotonic increasing function of $n$: if $X$ has $n$ equally likely outcomes and $Y$ has $m>n$ equally likely outcomes, then $Y$ has more uncertainty is than $X$.
	\item The value of H should not depend on how we structure the process generating the observable distribution. For example, if we have a random variable $X\in \{1,2,3,4\}$, we can model a measurement of $X$ as resulting from a single chance event with four possible outcomes, or as a sequence of two dependent events, each with two possibilities (conditional on the first outcome). If we are not allowed to look at the intermediate process, then the entropy of $X$ should be the same for each model.
\end{enumerate}

Due to the intuitively appealing properties of entropy noted by Shannon and others, both Shannon Entropy and Differential Shannon Entropy have been recommended as tools for developing uncertainty distributions (\cite{Jaynes57}, \cite{Jaynes82}). Maximum-entropy (ME) distributions are defined by their constraints, with several common examples given below:

\begin{itemize}
	\item $X\sim \textrm{Uniform}$ if $X \in [a,b],-\infty<a < b< \infty$ 
	\item $X\sim \textrm{Beta}$ if $X\in [a,b], -\infty<a < b< \infty, E[X]=\mu, V[X]=\sigma^2$
	\item  $X\sim \textrm{Exponential}$ if $X \in [0,\infty), E[X]=\mu$
	\item $X\sim \textrm{Normal}$ if $X\in (-\infty,\infty), E[X]=\mu, V[X]=\sigma^2$
\end{itemize}

The fact that the normal distribution emerges as a ME distribution is quite a satisfying result from a statistical and mathematical standpoint, as it concords with the ubiquity of this distribution and provides support for its use as a "minimally informative" choice in modeling.

However, while entropy as a measure of \textit{information content} is well-established, its applicability to modeling uncertainty is less clear cut. In particular, \cite{Uffink96} notes that the Shannon Entropy is just one of many "entropy-like" measures out there. Other examples are distributions maximizing the class of R\'{e}nyi-entropies, of which Shannon Entropy is a special case, or distributions that are invariant under a pre-specified transformation. What is interesting is that these forms of "entropy" also satisfy very plausible assumptions about how a measure of information should behave, although the details of these assumptions differ among the models (\cite{Uffink95}). Finally, even within the ME paradigm, the form of the resulting entropy maximizer depends critically on the form of the constraints (\cite{Uffink95}), implying that the MEP does not remove subjectivity as much as push it down one level.

In addition, from a decision-theoretic viewpoint, it is not clear how a ME distribution is affecting the decisions. Is it making us more conservative, less conservative? Are we emphasizing particular outcomes by maximizing the entropy? If so, why are those outcomes justified in being given more weight, and in particular the specific weights prescribed by the MEP? The above questions are not intended to be a pointed critique, and certainly not any form of rebuttal, of the MEP or the resulting distributions. Instead, the intent is merely to highlight the ambiguities inherent in attempting to model ignorance as opposed to information. The ambiguity is exacerbated by the existence of alternative principles of similar intuitiveness or level of justification (e.g., Jeffrey's invariance principle, R\'{e}nyi-entropies).

The novel approach presented in this paper no doubt also possesses its own set of ambiguities and unanswered questions. However, like other methods for modeling uncertainty, it has several features that may appeal to  various practitioners and/or be particularly applicable in certain contexts.

\section{Uniformity}
\label{sec:unif}

The approach taken in this paper seeks to maximize the \textit{uniformity} of a probability distribution, where "uniform" is meant in the sense of the "equal probability" of a random variable for various outcomes. In particular, this paper focuses on maximizing the uniformity of bounded, continuous-valued random variables with smooth distribution functions (i.e., the cumulative distribution function (CDF) is differentiable over the entire domain). Like other measures of uncertainty, \textit{uniformity} is developed from a base of intuitively plausible presuppositions and then extended to less familiar domains. In particular, any uniformity measure for a continuous, bounded random variable should possess two very basic properties:

\begin{enumerate}
	\item A degenerate distribution (i.e., step function CDF) is the least uniform. 
	\item The uniform distribution (not surprisingly) is the most uniform distribution. 
\end{enumerate}

Given a measure of uniformity, we formulate the central "principle" for choosing a distribution using this measure:

\subparagraph*{Principle of Maximum Uniformity:}
Of all continuous, finite-domain distribution functions  that are consistent with the available information, choose the one with the highest measure of uniformity.\\

The rationale for this principle is that we want to choose a distribution "as uniform as possible" from among the available distributions, so that we are not, \textit{a priori}, overemphasizing any regions of the sample space beyond what is necessary to be consistent with our prior assumptions or available information.

To actually implement this principle, we need to quantify how close a given distribution is to a uniform distribution. The following observation will prove instrumental to the definition of such a quantity: the set of points $\{(x,F_U(x)):x\in[a,b]\}$ generated by the CDF of the uniform density function, $F_U$, on $[a,b]$ takes the shortest possible path from $(x_0,y_0)=(a,0)$ to $(x_1,y_1)=(b,1)$. This is a defining feature of the uniform CDF (and of straight lines in general). Hence, we used the path length of the CDF as the basis for our quantity:

\subparagraph{Definition: Uniformity Index}
Given a continuous random variable $X\in [a,b]$, with probability density function $f_X$, the \textit{uniformity index} of $f_X$ is defined as:\\ \[\mathcal{U}(f_X):= \frac{\sqrt{1+(b-a)^2}}{\int_a^b \sqrt{1+[f_X(t)]^2}dt}\]

\paragraph{} The uniformity index is simply the ratio of the path length of the uniform CDF to the path length of the CDF of X ($F_X$) on $[a,b]$. It satisfies the two basic properties of a uniformity measure listed earlier: the longest path is the step function CDF associated with the \textit{degenerate} density function $\delta_a(x)$ and the shortest path is produced by the uniform CDF on $[a,b]$. In general, $\frac{1}{\sqrt{2}} \approx 0.707 \leq \frac{\sqrt{1+(b-a)^2}}{1+b-a} \leq \mathcal{U}(f_X) \leq 1$. Note that for a given domain $[a,b]$, the uniformity index varies only as a function of the path length of $f_X$; hence, maximizing the uniformity is equivalent to finding the \textit{Shortest Path Distribution} (SPD)  that satisfies a given set of constraints $C$.

\section{SPD under Raw Moment Constraints}

Given a domain $[a,b]$ and set of constraints $C$, we are seeking to minimize the path length $L[F]$ of the cumulative distribution function $F$ defined on $[a,b]$, subject to a set of constraints $C$ (i.e., $F$ is part of a family of distributions $\mathcal{P}_C$ satisfying $C$):\\
\[  \argmin_{F\in\mathcal{P}_C}L[F]\equiv\argmin_{F\in\mathcal{P}_C}\int_a^b \sqrt{1+\left(\frac{dF}{dx}\right)^2}  dx\]
\\
This type of problem is typical in the \textit{Calculus of Variations}, where we are seeking to find a univariate function $f(x)$ that minimizes a functional $I[f]\equiv \int_a^b G(x,f,f')dx$ . Candidates for optimal functions are identified as solutions to the \textit{Euler-Lagrange} equation:

\[ \frac{\partial G}{\partial f} - \frac{d}{dx}\left(\frac{\partial G}{\partial f'}\right) =0 \]

In general, the Euler-Lagrange equation is merely a \textit{necessary} condition for a minimum. Second-order conditions are required to verify that $f$ is indeed a minimum and not a maximum or merely a stationary point (think "plateaus" of some cubic polynomials). A necessary condition for $f$ being a minimum of $I[f]$ is the \textit{Legendre Condition}:

\[ \frac{\partial^2G}{\partial (f')^2}  \geq 0 \;\mathrm{for}\; x\in [a,b] \]

This paper will focus on solving the above problem for the path length functional $L[F]$ under constraints on the \textit{raw} moments of $f=F'$ and a basic non-negativity constraint on the probability density function $f$ (i.e., $f(x)\geq 0\;\forall x \in [a,b])$. The functions associated with the $SPD$ constraints on all raw moments up to $m$ is given by the vector-valued function $C^m$:

\begin{equation}
 C^m(f;\vec{\mu}):=\left(\int_a^b x^if(x)dx-\mu_i\right)_{i=1}^m,\;\mathrm{where}\; \vec{\mu}=(\mu_0,\mu_1,...,\mu_m)\;\textrm{ and }\;\mu_0\equiv1
 \label{eqn:C}
\end{equation}

The constraints in this problem are referred to as \textit{isoperimetric} constraints (alluding to the motivating problem of finding a shape that maximizes the enclosed area given a fixed \textit{perimeter} length). As with classical constrained optimization, constraints can be incorporated using \textit{Lagrange Multipliers}, $\lambda_i$, which turn a constrained problem into an unconstrained problem. In particular, the integrand for isoperimetric constraint $i$ is multiplied by $\lambda_i$ and \textit{added} to the integrand of $L[F]$ to form the \textit{Lagrangian}, $\phi$, which will be used to find the optimal function:

\begin{equation}
 \phi[x;f,\lambda] \equiv \sqrt{1+f(x)^2} + \sum_{i=0}^m \lambda_ix^if(x),\;\mathrm{ where } \;f(x)=F'(x)
 \label{eqn:Lag}
\end{equation}  

Applying the Euler-Lagrange equation to $\phi$, we can derive the general form of a solution:

\[\frac{\partial \phi}{\partial f} - \frac{d}{dx}\left(\frac{\partial \phi}{\partial f'}\right) = \frac{f}{\sqrt{1+(f)^2}}+ \sum_{i=0}^m \lambda_ix^i =0\implies\]
\begin{equation}
f(x;\lambda)=\frac{d}{dx}F(x;\lambda)=\frac{\sum_{i=0}^m\lambda_ix^i}{\sqrt{1-\left(\sum_{i=0}^m\lambda_ix^i\right)^2}}
\label{eqn:f} 
\end{equation} 

Where in the last equation, we took the positive root of the numerator, as the sign is inconsequential with the undetermined multipliers. We will now check the  Legendre condition for $\phi$:

\[ \frac{\partial^2\phi}{\partial (f')^2} = \frac{\partial^2}{\partial (f')^2} \left( \sqrt{1+f(x)^2} + \sum_{i=0}^m \lambda_ix^if(x)\right)=0  \]

This result indicates that our solution is not ruled out as a  minimum. We can informally check the actual solutions by comparing them to a trial solution. If the solution has a longer path length than our trial solution, we know its not a minimum. However, if it has a shorter length, then we are likely looking at a minimum. This is the approach we will use on the specific solutions derived in this paper.

We will now need to solve the system of integral constraints for the vector of Lagrange multipliers $\mathbf{\lambda}:=(\lambda_0,\lambda_1,...,\lambda_n)$, such that the \eqref{eqn:f} is non-negative and satisfies the isoperimetric constraints \eqref{eqn:C} : 

\[\mathrm{Find }\;\mathbf{\lambda}: \left\{\frac{\sum_{i=0}^m\lambda_ix^i}{\sqrt{1-\left(\sum_{i=0}^m\lambda_ix^i\right)^2}}\geq 0,\; \forall x\in[a,b];\;\;C^m\left(\frac{\sum_{i=0}^m\lambda_ix^i}{\sqrt{1-\left(\sum_{i=0}^m\lambda_ix^i\right)^2}};\vec{\mu}\right)=\mathbf{0}\right\}\]

\section{Numerical Solution}

The form of the density function precludes a simple analytical solution for $\lambda$. Instead, we numerically estimated $\lambda$ using a piecewise uniform approximation $f^*(x;\lambda)$ of the true density function $f(x;\lambda)$. First, the domain of $f$ was broken into $n$ equal-length intervals $\Delta_i=[x_{i-1},x_i)=[a+(i-1)\delta_n,a+i\delta_n],\;i=1...n$, where $\delta_n=\frac{b-a}{n}$. Each interval has a midpoint, $p_i=\frac{a+(i-1)\delta_n+a+i\delta_n}{2}=\frac{2a+(2i-1)\delta_n}{2}$, which is used for calculating the value of the density over that interval. Using this approach, we were able to replace the isoperimetric constraints (which are expressed as integrals) with Riemann sums over the partition $\{\Delta_i\}$:

\[ C^m_k= \int_a^b x^kf(x)dx -\mu_k \approx \int_a^b x^kf^*(x)dx -\mu_k=\sum_{i =1}^n \left(\int_{\Delta_i} f(p_i)x^k dx\right) -\mu_k= \]
  \begin{equation}
  \sum_{i=1}^n f(p_i;\lambda)\frac{x_i^{k+1}-x_{i-1}^{k+1}}{k+1} -\mu_k= \langle\mathbf{f}(\mathbf{p};\mathbf{\lambda}),\mathbf{\delta}^k_n\rangle-\mu_k:= M_n^k(\lambda)
  \label{eqn:M}
  \end{equation} 
\[\mathrm{where}\; \delta^k_{n,i}:= \int_{x_{i-1}}^{x_i} z^k dz \]

Each summand on the left hand side of \eqref{eqn:M} represents the contribution of the interval $\Delta_i$ to the $k^{th}$ raw moment of the piecewise uniform approximation to $f$. Note that this reduces to the simple Riemann sum over $f$ with partition length $\delta_n$ in the case of the "zeroth" moment ($k=0$). For a given partition size $n$, we can estimate $\lambda$ using a least squares solution to the constraints of the approximate problem:

\begin{equation}
\argmin_{\lambda} \sum_{j=1}^m [M_n^j(\lambda)]^2
\label{eqn:obj} 
\end{equation}

Theoretically, this is all we need. However, from a numerical perspective, the form of \eqref{eqn:f} as a function of $\lambda$ has two problems :

\begin{enumerate}
	\item The numerator can become negative
	\item The denominator can become complex
\end{enumerate}

Therefore, we needed to define a feasible region for $\lambda$ to ensure we got physically sensible answers. This was be accomplished using the following constraint set on $\lambda$:

\begin{equation}
0\leq \sum_{i=0}^m\lambda_ip_j^i < 1,\;j=1...n
\label{eqn:barrier}
\end{equation}

Conveniently, \eqref{eqn:barrier} defines two linear half-spaces, which defines a convex polytope in $\mathbb{R}^m$.\\

We solved the above formulation using the R statistical programming language to run the \textit{auglag} nonlinear optimization routine in the \textit{alabama}  package, with Nelder-Mead as the selected optimization method. The algorithm converged in some specific instances; however, numerical tests indicated that directly using the parameterized form of the density $f(p_i;\lambda)$ in the optimization model induced instability and poor convergence for some values of the constraints. Therefore, we developed a second approach, which took a more direct, high-dimensional optimization route.\\

In the second approach, we used the same set of partitions $\{\Delta_i\}$ and midpoints $p_i$ as in the first approach, but we treated the density values at each midpoint (i.e., $f(p_i)$) as a decision variable. This change resulted in an $n$ variable minimization problem in $f(p_i):=f_i$, where we are approximating the true density function $f(x;\lambda)$ by a piecewise constant function $f^*(x)$ that takes value $f_i$ if $x \in \Delta_i$. If a smooth, continuous solution exists to the true optimization problem, then the sequence of approximate solutions will also converge (pointwise): $\lim\limits_{n\to \infty} \sum_{i=1}^n \mathbf{1}(x)_{x\in \Delta_i}f_i(x) = f(x)\;\forall x\in[a,b]$, so $f^*(x) \to f(x) \;\mathrm{pointwise}$. The new optimization model becomes:

\begin{equation}
 \argmin_{f_i} \delta_n\sum_{i=1}^n \sqrt{(1+f_i^2)}\;\;\; \mathrm{s.t.}
\; M_n^k=0,\;0\leq k\leq m;\\  
 f_i\geq 0,\;i=1...n 
\end{equation}

Where $M_n^k := \langle\mathbf{f},\mathbf{\delta}^k_n\rangle-\mu_k$ as in \eqref{eqn:M}, but now the vector of density values are the actual decision variables, not values parameterized over $\lambda$. This more direct approach (using R with $auglag$ and the $nlminb$ optimization routine) had better convergence properties for the problems we investigated in this paper, with first and second order KKT conditions being consistently satisfied. However, we felt that the analytical solution to the Euler-Lagrange equation still has theoretical and conceptual value, despite not leading to the best numerical approach.

The next section will show the typical shape of the resulting SPDs under low-order moment constraints. Each of these will be graphically compared to their equivalent ME distribution and the CDF lengths will be calculated for each. The paper will end with a discussion of the key insights from this investigation.

\section{Typical solutions for $m\in \{0,1,2\}$}
\subsection{Uniform $(m=0)$}
This is the simplest case, where we only require that the probability equal 1. The resulting SPD is a uniform distribution, which is identical to the ME distribution. The uniform distribution is trivially the shortest path distribution under these conditions.

\subsection{Exponential $(m=1)$}
We start to see some more structure when we also constrain the mean. In this case, we are searching for a SPD on $[0,0.1]$ with mean $0.04$. The equivalent ME distribution is the truncated exponential distribution $TEXP(\lambda,b)$, where $\lambda=12.3,b=0.1$. (Fig. 1) The path length for the truncated exponential solution is $1.017$, whereas the path length of the SPD distribution is $1.006$, verifying that indeed the SPD distribution has a shorter path length. 

While absolute path-length differences appear slight, this is primarily due to an issue of scale. To correct for this, we calculated the ratio of each distribution's respective difference from the CDF path length of a uniform distribution on $[0,.01]$ (i.e., this would be $\sqrt{1+(0.1)^2}=1.005$), to get a "difference ratio" ($r$) of $11$. The general formula is shown below:

\begin{equation}
 \textrm{Difference Ratio}(f_{ME},f_{SP}):=\frac{L[f_{ME}]-\sqrt{1+(b-a)^2}}{L[f_{SP}]-\sqrt{1+(b-a)^2}}
 \label{eqn:diffratio}
\end{equation}\\ 

In this case, a difference ratio of $11$ means that the SPD CDF path length is $11$ times closer to the straight-line distance (shortest possible path) than the CDF path length of the associated truncated exponential distribution. Essentially, we are using the path length of the uniform CDF over $[0,.01]$ as the point of reference, as $0$ is not a feasible path length. \\

\begin{figure}[H]
	\centering
	\includegraphics[width=0.7\linewidth]{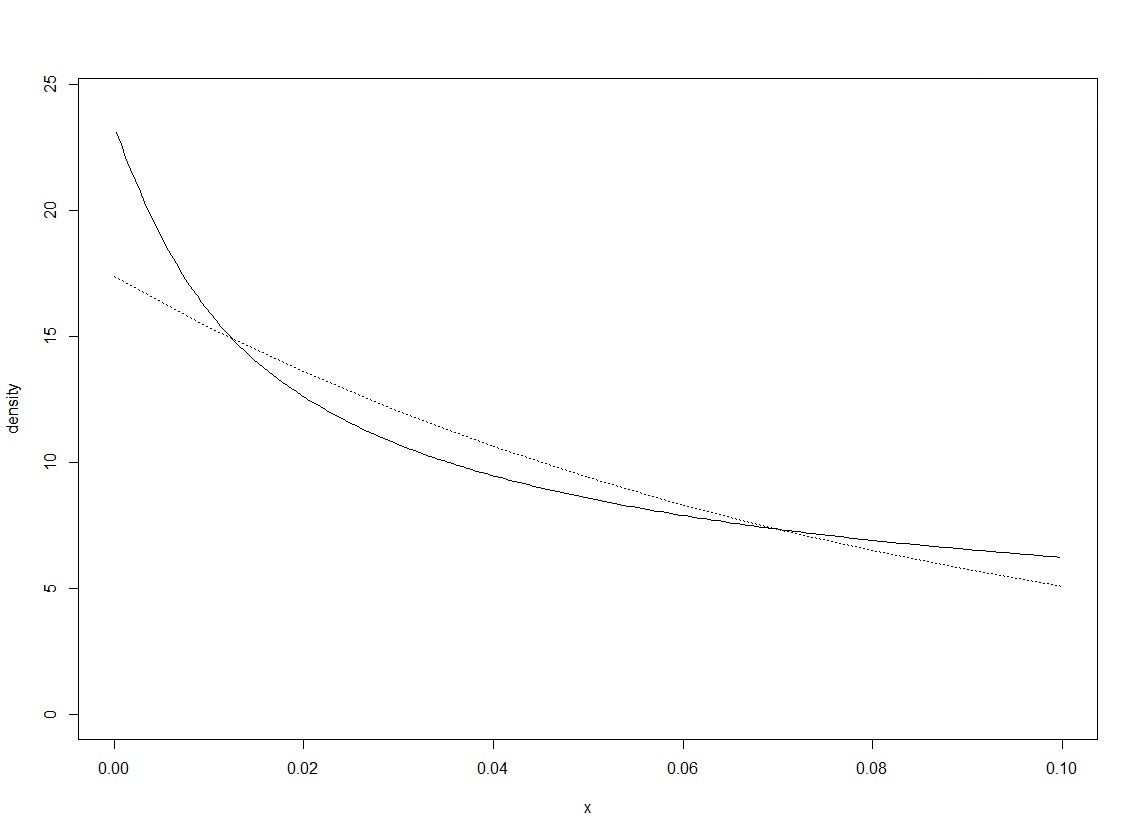}
	\caption[Figure 1]{SPD (solid line) and ME distribution (dotted line) on $[0,0.1]$  for $m_1=0.04$}
	\label{fig:Fig1}
\end{figure}

As an additional exploration, we also examined how the solutions change as the upper bound $b$ of the random variable is increased. 

\begin{figure}[H]
\centering
\includegraphics[width=0.7\linewidth]{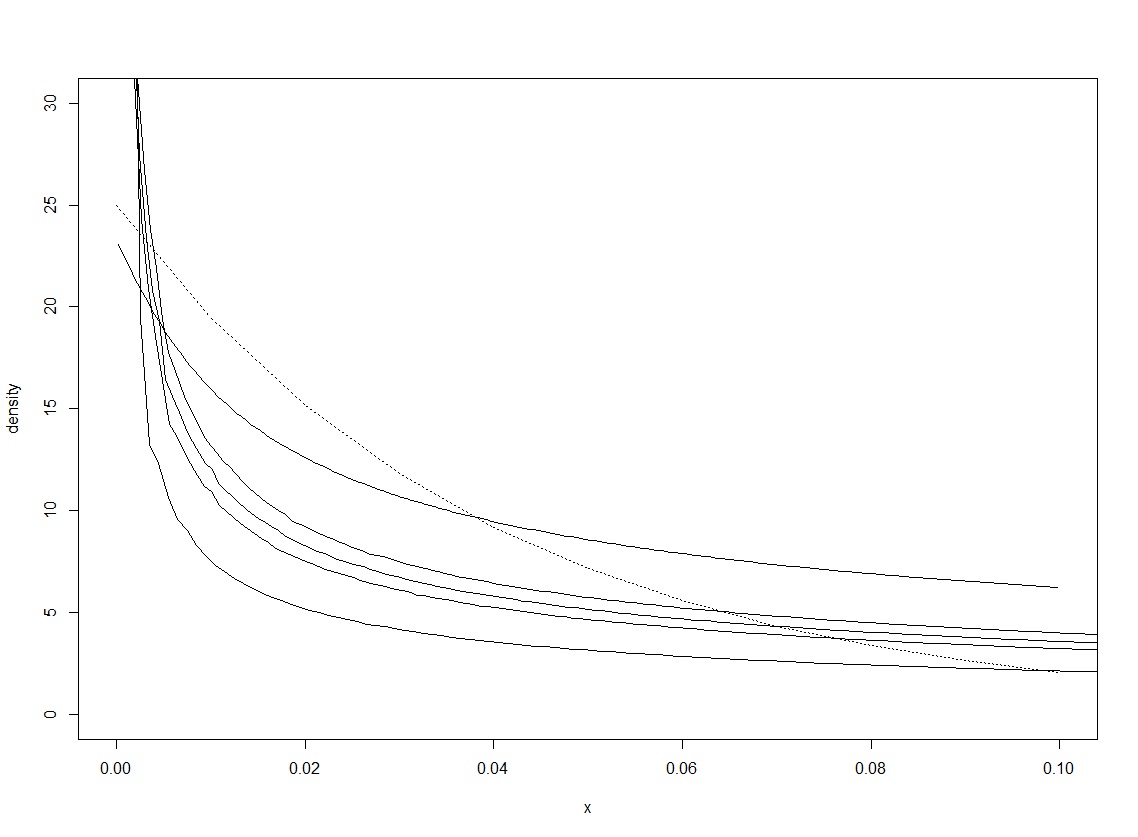}
\caption[Figure 2:]{Sequence of SPDs for $m_1=0.04$ with increasing upper bounds (solid lines) and ME distribution for unbounded case (dotted line).}
\label{fig:Fig2}
\end{figure}

From Figure \ref{fig:Fig2} it appears that qualitatively the SPD becomes increasingly more peaked, suggesting that $\lim\limits_{b\to \infty}SPD(m_1=0.04) \;\dot{=}\; \delta(0)$. However, $E[\delta(0)]=0\neq 0.04$, which would imply that there is no unbounded SPD (consistent with the fact that the sequence path lengths of the associated CDFs is unbounded).

\subsection{Bell and Bowl $(m=2)$}
\begin{figure}[H]
\centering
\includegraphics[width=0.7\linewidth]{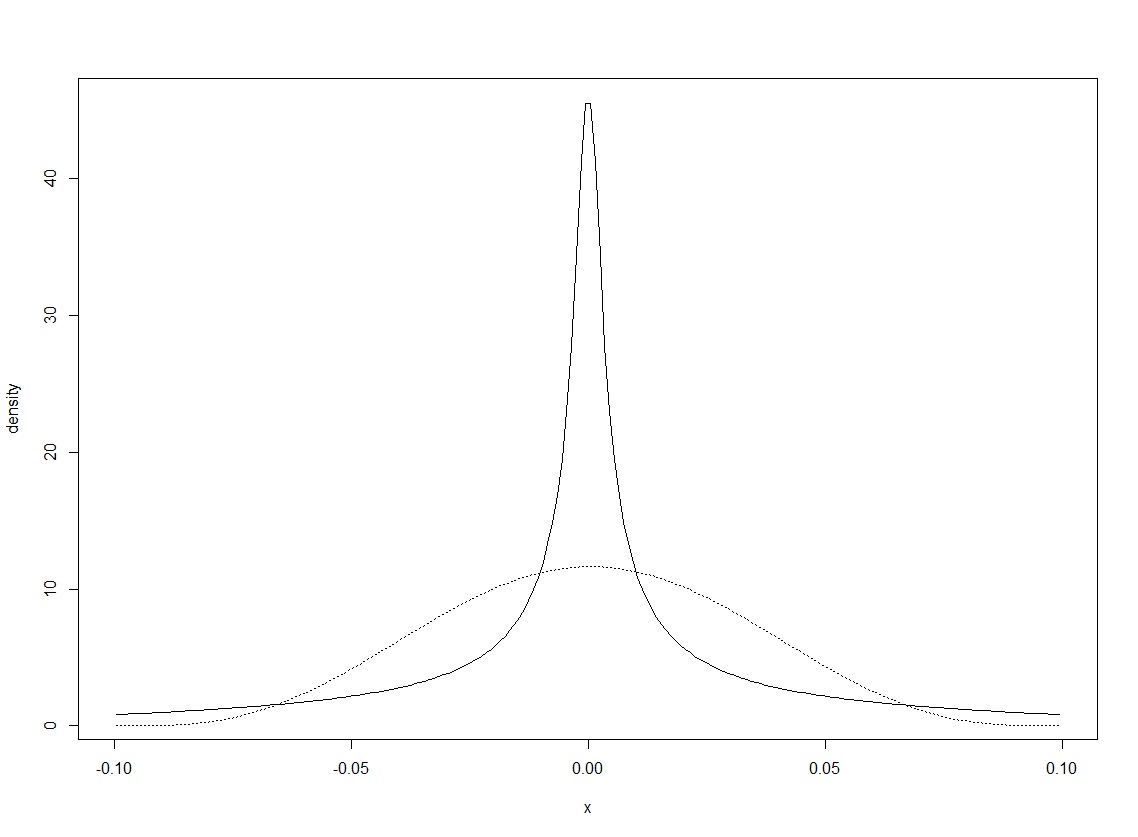}
\caption[Figure 3:]{SPD for $m_1=0,m_2=0.001$ on $[-0.1,0.1]$ (solid line) and Bell Beta (dotted line)}
\label{fig:Fig3}
\end{figure}

\begin{figure}[H]
	\centering
	\includegraphics[width=0.7\linewidth]{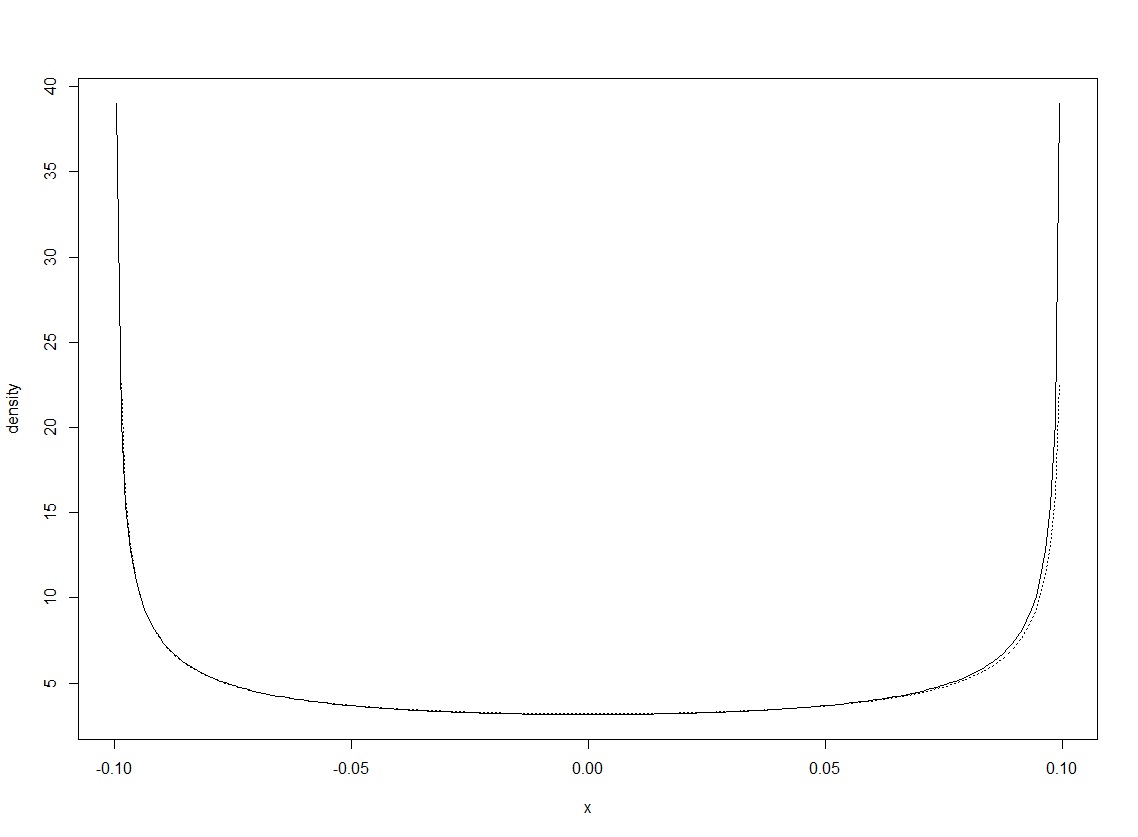}
	\caption[Figure 3:]{SPD for $m_1=0,m_2=0.005$ on $[-0.1,0.1]$ (solid line) and  Bowl Beta (dotted line)}
	\label{fig:Fig4}
\end{figure}

This was the most complex case examined. We calculated two SPDs on $[-0.1,0.1]$ with mean $0$ and second-moments (here:\textit{variance}) equal to $0.001$ and $0.005$, respectively. The equivalent ME distributions are Beta distributions on $[-0.1,0.1]$ with parameters $(\alpha_1=\beta_1=4.5)$ and $(\alpha_2=\beta_2=0.5)$ , respectively. We called the first case the "Bell" and the second case the "Bowl" Beta distributions to indicate their qualitative shapes. We found that the SPD with the same first and second moments as the Bell Beta distribution was markedly more peaked (Figure \ref{fig:Fig3}), while the Bowl Beta distribution and moment-matched SPD were almost identical (Figure \ref{fig:Fig4}). \\

\begin{figure}
\centering
\includegraphics[width=0.7\linewidth]{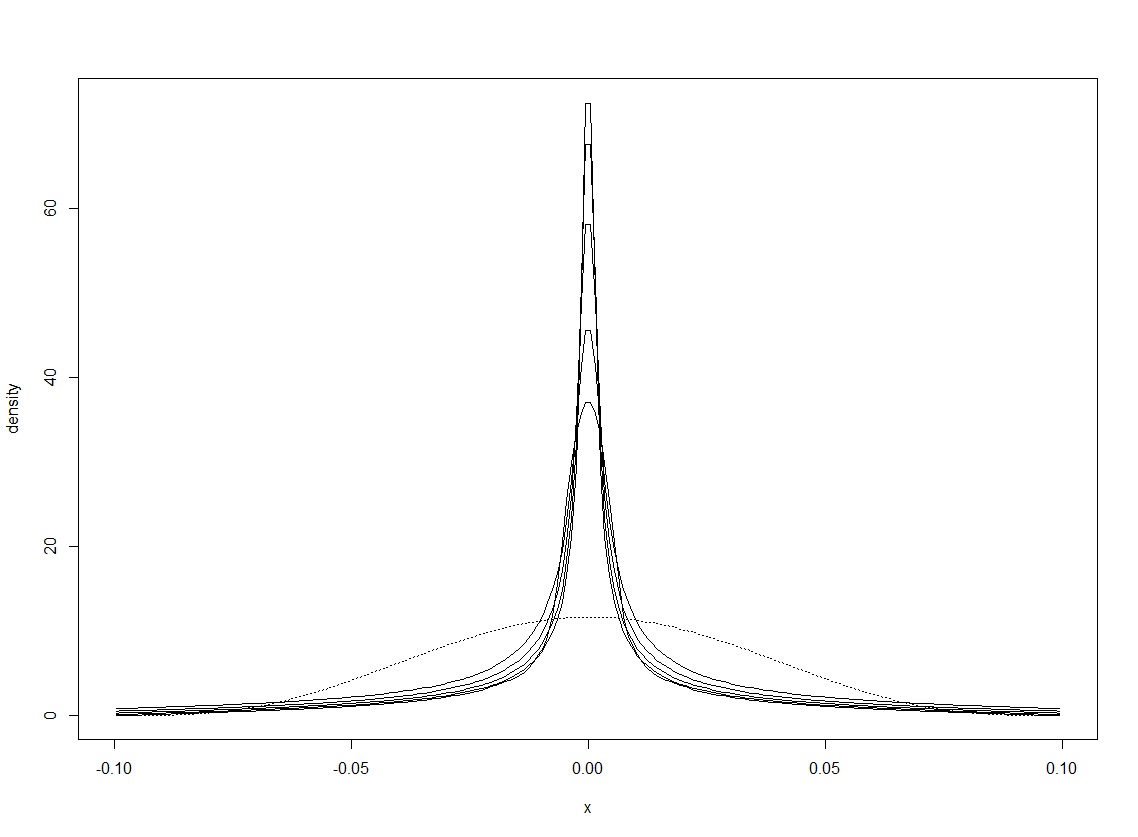}
\caption[Figure 5:]{Sequence of SPDs $(m_1=0,m_2=0.001)$ with symmetrically increasing support (solid lines). Moment-matched normal distribution with unbounded support (dotted line).}
\label{fig:Fig5}
\end{figure}

The path length was $1.06$ for the Bell Beta CDF and $1.04$ for the moment-matched SPD CDF, yielding a difference ratio of $2$.  The Bowl Beta and moment-matched SPD CDFs had the same path length of $1.02$. Similar to the case with $m=1$, we examined the sequence of Bell-shaped SPDs with increasing support (Figure \ref{fig:Fig5}) and compared them to their moment-matched normal distribution. 

As shown in Figure \ref{fig:Fig5}, it appears that $\lim\limits_{b\to \infty} SPD(m_1=0,m_2=0.001) \;\dot{=}\; \delta(0)$, similar to the fixed-mean SPD in the previous section. Again, this suggests that the SPDs must be bounded.

\section{Discussion}

We see that the SPDs do not, as a general rule, match the results given by the Maximum Entropy Principle. Therefore, how can we interpret the difference between an SPD and its associated ME distribution? In broad terms, it appears that SPDs emphasize "fragile stability" while the ME distributions exhibit "stable instability". A process operating according to a ME distribution will rapidly "fill" its outcome space, reaching its equilibrium distribution quickly due to its high entropy. In contrast, a process operating according to an SPD will appear to be relatively stable and well contained, only to suddenly produce an extreme outlier (relative to the central 95$\%$ of the distribution).\\

The punctuated/catastrophic behavior of SPDs is reminiscent of the types of behavior discussed in Nassim Taleb's \textit{The Black Swan}, where Taleb argues that the variance between outcomes is a poor measure of risk if the underlying distribution has high kurtosis (thus high tail-risk). It is interesting that taking a very geometric/literal approach to uniformity produces such different densities when compared to distributions based on maximizing an abstract \textit{measure}.  The generality of this behavior is an open question: does it arise only with path length, or does it arise whenever one is optimizing a \textit{metric} as opposed to a \textit{measure} as a function of an input distribution. Regardless, the results of this analysis point out that it is far from clear how "uncertainty" is best quantified, with different, plausible approaches yielding very different results. By emphasizing extreme behavior, SPDs provide a novel perspective on the relationship between risk, ignorance, probability, and uncertainty

\bibliographystyle{Chicago}
\bibliography{SPDLaTeX}

\end{document}